%Paper: hep-th/9212144
%From: RBROOKS@IRENE.MIT.EDU (Roger Brooks)
%Date: Wed, 23 Dec 1992 12:16:35 -0500 (EST)

% MACROS
%
\hsize=13cm
\hfuzz=20pt
\vsize=18cm
\tolerance=10000
\magnification=1200
%\hoffset=2.5truecm
%\voffset=2.5truecm
%
% FONTS

\font\headfont=cmbx10 scaled 1440
\font\namefont=cmr10
\font\initialfont=cmr10 scaled 1200
\font\addfont=cmti10
\font\fntefont= cmr7 scaled 1200

\def\fracfont#1{{\the\scriptfont0 #1}}
%\font\fracfont=cmr7

% Macros used in math mode

\def\sp{\ }
\def\seq{\sp=\sp}
\def\pls{\sp+\sp}
\def\mi{\sp-\sp}
\def\pd{\partial}
\def\mbox#1#2{\vcenter{\hrule width#1in\hbox{\vrule height#2in
   \hskip#1in\vrule height#2in}\hrule width#1in}}

\def\frac#1#2{{\fracfont{{#1}\over{#2}}}}
\def\frc#1#2{\leavevmode\kern .1em
             \raise .5ex\hbox{\the\scriptfont0 $#1$}\kern -.1em /
             \kern-.15em\lower .25ex\hbox{\the\scriptfont0 $#2$}}
\def\half{\frac{1}{2}}
\def\inv#1{\frac{1}{#1}}
\def\oplu#1{\leavevmode\raise .3em\hbox{$\oplus$}
            \kern-1.51em\lower.4em\hbox{\the\scriptfont0 #1}}
\def\nrmord#1{\leavevmode\raise .3em\hbox{\the\scriptfont0 $#1$}
            \kern-.43em\lower.3em\hbox{\the\scriptfont0 $#1$}}

\def\slash#1{\hbox{$#1$\kern-.52em\hbox{$/$}}}

%
% Macros used in text
%
\def\drawline#1{{\hbox to #1truein{\hrulefill}}}

\def\fnote#1#2{\baselineskip 14pt{\footnote{$^#1$}{\fntefont #2}}
               \baselineskip 18pt}
\count101=0
\def\fnoter#1{\advance\count101 by 1
              \baselineskip 14pt{\footnote{$^{\backslash \number\count101
              }$}
                                            {\fntefont #1}}
               \baselineskip 18pt}
%
% Title page macros
%
\def\title#1#2#3{\centerline{\bf{\headfont #1}}
			\medskip
			\centerline{\bf{\headfont #2}}
			\medskip
			\centerline{\bf{\headfont #3}}
		       }
\def\contract{$\!\!\!$
	\fnote{\star}{This work is supported in part by funds provided by the
		      U. S. Department of Energy (D.O.E.) under contract
		      \#DE-AC02-76ER03069.
		     }
	     }
\def\author#1#2#3#4#5{\vskip 0.5truein
	\centerline{{\initialfont #1}{\namefont #2}{\initialfont #3}
		    {\initialfont#4}{\namefont#5}}
		     }
\def\address{\medskip
	\centerline{\addfont{Center for Theoretical Physics,}}
	\centerline{\addfont{Laboratory for Nuclear Science}}
	\centerline{\addfont{and Department of Physics}}
	\centerline{\addfont{Massachusetts Institute of Technology}}
	\centerline{\addfont{Cambridge, Massachusetts 02139 U.S.A.}}
	    }
\def\abstract{\vskip 0.75truein
		\baselineskip 18pt plus 2pt minus 2pt
		\centerline{\bf ABSTRACT}
		\medskip
		\par
		}
\def\ctp_no#1#2{\noindent CTP \#{#1}\hfill #2}
\def\hep_no#1{\leftline{#1}}
%
% Acknowlegdments
%

%
% Reference Macros
%
\def\refs{\vfill\eject
	  \centerline{\bf REFERENCES}
	  \vskip 0.9truecm}
\def\ref#1#2{\item{#1.}{#2.}\smallskip\goodbreak}

\def\np#1{{\it Nucl. Phys.} {\bf B#1}}
\def\cmp#1{{\it Commun. Math. Phys.} {\bf #1}}

\def\cqg#1{{\it Class. Quan. Grav.} {\bf #1}}

\def\pr#1{{\it Phys. Rev.} {\bf #1}}
\def\pl#1{{\it Phys. Lett.} {\bf #1B}}

% Greek letters
%
\def\a{\alpha}
\def\b{\beta}

\def\d{\delta}

\def\e{\epsilon}
\def\g{\gamma}

\def\l{\lambda}

\def\th{\theta}

\def\z{\zeta}
%
% Caligriphic letters
%

\def\cg{{\cal G}}
\def\ch{{\cal H}}

\def\cp{{\cal P}}

\def\RefGrossMende{1}
\def\RefWittenTQFT{2}
\def\RefBrooksSpec{3}
\def\RefAchucarro{4}
\def\RefWittenCSG{5}
\def\RefChamseddine{6}
\def\RefChamWyl{7}
\def\RefAsh{8}
\def\RefBengtson{9}
\def\RefRomano{10}
\def\RefBirminghamRev{11}
\def\RefHorowitz{12}
\def\RefBFV1{13}
\def\EqPSI{1}
\def\EqETA{2}
\def\EqQ{3}
\def\EqQTRANS{4}
\def\EqHQQBAR{5}
\def\EqH{6}
\def\EqSBF{7}
\def\EqCons{8}
\def\EqETC{9}
\def\EqCA{10}
\def\EqIH{11}
\def\EqFDIFF{12}
\def\EqDIFF{13}
\def\EqHthree{14}
\def\EqQthree{15}
\def\EqHtwo{16}
\baselineskip 15pt
\def\LG{\hbox{{\rm L} \kern -0.7em {\rm G}}}
\def\IH{\hbox{{\rm I} \kern -0.55em {\rm H}}}
\def\IR{\hbox{{\rm I}\kern -0.2em {\rm R}}}
\def\dual#1{\kern .2em ^*\kern -.2em #1}
\def\dsig{\d_\Sigma(\vec{x}-\vec{y})}
\def\ie{{\it i.e.}}
\def\pd{\partial}

\def\diffs{{\rm Diff}(\Sigma)}
\def\etab{\bar\eta}
\def\thb{\bar \th}
\def\zb{\bar \z}
\def\image{{\rm Im}}
\def\ihp{\IH_\perp}
\def\lrpd{\leavevmode\raise .6em\hbox{$\leftrightarrow$}
            \kern-.85em\hbox{$\pd$}}
\def\rhob{{\bar \rho}}
\def\xib{{\bar \xi}}
\def\sigmaa{\leavevmode\raise .15em\hbox{$\sum$}}
%%%%%%%%%%%%%%%%%%%%%
% PAPER STARTS HERE %
%%%%%%%%%%%%%%%%%%%%%
\title{Diff($\sigmaa$) and Metrics from Hamiltonian--TQFT's}
	{in 2+1 Dimensions\contract}
	{}

\author{R}{OGER}{}{B}{ROOKS}

\address

\abstract
The constraints of $BF$ topological gauge theories are used to construct
Hamiltonians which are anti-commutators of the BRST and anti-BRST operators.
Such Hamiltonians are a signature of Topological Quantum Field Theories
(TQFT's). By construction, both classes of topological field theories share
the same phase spaces and constraints. We find that, for 2+1 and 1+1
dimensional space-times foliated as $M=\Sigma\times\IR$, a homomorphism exists
between the constraint algebras of our TQFT and those of canonical gravity.
The metrics on the two-dimensional hypersurfaces are also obtained.
\vskip 4truecm
\ctp_no{2175}{December 1992}
\hep_no{hep-th/yymmnn}
\vfill\eject

A popular notion in the literature on quantum gravity is
that of an unbroken phase of general covariance in which
the expectation value of the metric is zero; \ie, the metric
is degenerate.  Although string theory [\RefGrossMende]
has provided input into our comprehension of such a phase,
a complete picture is far from emerging.  Progress
in this direction was made with the construction of a class
of Topological Field Theories (TFT's) known as Topological Quantum
Field Theories (TQFT's) [\RefWittenTQFT].
These were originally intended to be used as a machinery
for computing topological invariants.  Crucial to these theories are
topological/BRST
symmetries which obviate dynamical degrees of freedom; indeed, the
classical action is topological.
Nevertheless, physicists have found these to be attractive
as they are suggestive of the unbroken phase of general covariance.
In fact, it was hoped that we might be able to liberate the metric
via the breaking of these symmetries.
However, on general grounds, it is not
possible to either spontaneously or dynamically break these
symmetries [\RefBrooksSpec].  Evidently, in order to
arrive at the desired result, either a new means of symmetry
breaking must be found or we must alter our approach to TQFT's.
We might search for a method for identifying the diffeomorphism
invariance constraints of General Relativity (GR) in TQFT's.
As there is no dynamical metric in TQFT's, it has thus far been
unclear how to derive these constraints.

Almost in conjunction with the development of TQFT's, a different
class of
topological actions for gravity in low dimensions ($D=2,3$) have
appeared in the literature [\RefAchucarro-\RefRomano].  These are
complete theories of pure, low dimensional gravity; in both the
broken and unbroken phases.
They exist by virtue of the fact that gravity in
low dimensions is non-dynamical. For example, the graviton and
graviton-dilaton systems have no degrees of freedom in 2+1 and
1+1 dimensions, respectively.  This second class of
TFT's are known as $BF$ topological gauge
theories [\RefBirminghamRev]. The $B$ field is a Lagrange multiplier
which forces the field strength, $F$, to vanish.  When these
theories are applied to gravity, $F$ is identified with the
Riemann curvature tensor. Hence the $BF$ theories are not applicable
as theories of gravity in four or higher dimensions. Since TQFT's
are generically not as constrained as $BF$ theories, they are better
candidates for the unbroken phase of general covariance in higher
dimensions. In this note, we will find that it is possible to
exploit the successes of the $BF$ gravity theories to obtain
metrics and a representations of diffeomorphism invariance in
Hamiltonian-TQFT's.  By ``Hamiltonian-TQFT'' we mean any theory whose
Hamiltonian is an anti-commutator of BRST and anti-BRST operators.
As this is known to be a signature only of TQFT's, we postulate the
existence of TQFT actions from which our Hamiltonians may be derived.
Thus, we will drop the prefix ``Hamiltonian'' in the following.

To be precise, for two and three
dimensional manifolds $M=\Sigma\times\IR$,
we will build the Hamiltonians of
TQFT's from the phase spaces of $BF$ theories.
By construction, the reduced phase spaces of the TQFT and $BF$
theories will be the same.
For 2+1 dimensions, the metric restricted to $\Sigma$ will be
identified through the solutions of the constraints of the TQFT.
We will also obtain this metric via the identification of a
homomorphism between the constraint algebra of the TQFT and that
of GR.

Our construction of the Hamiltonian may be applied to arbitrary
theories with first class, irreducible constraints, $\Psi_\a$,
defined
on a phase space with coordinates $(p,q)$.  Let the constraints be
Grassmann even (generalization to Grassmann odd is immediate) and
form a semi-simple Lie Algebra:
$$[\Psi_\a(x),\Psi_\b(y)]\seq iC_{\a\b}{}^\g\Psi_\g(x)
\dsig\ \ .\eqno(\EqPSI)$$
Following BFV [\RefBFV1] quantization we extend the phase
space to include a pair of Grassmann odd fields, the ghost $\eta^\a(x)$
and the anti-ghost $\etab^\a(x)$ obeying the equal time anti-commutator
$$\{\etab^\a(x),\eta_\b(y)\}\seq \d_\b{}^\a \dsig\ \ .\eqno(\EqETA)$$
Then the nilpotent BRST and anti-BRST charges are constructed as
$$\eqalign{
Q&\seq \int_\Sigma \big(\eta^\a\Psi_\a \mi \half \etab^\a
[\eta,\eta]_\a\big)\ \ ,\cr
{\bar Q}&\seq\int_\Sigma \big(\etab^\a\Psi_\a \mi \half \eta^\a
[\etab,\etab]_\a\big)\ \ ,\cr}\eqno(\EqQ)$$
for which
$$\eqalign{
Q^2\seq&{\bar Q}^2\seq 0\ \ ,\cr
\{Q,\eta^\a\}\seq& -\half [\eta,\eta]^\a\ \ ,\cr
\{Q,\etab^\a\}\seq& \Psi^\a - [\eta,\etab]^\a\ \ ,\cr
[Q,\Psi_\a]\seq&  [\Psi,\eta]_\a\ \ ,\cr
\{{\bar Q},\eta^\a\}\seq& \Psi^\a  - [\etab,\eta]^\a\ \ ,\cr
\{{\bar Q},\etab^\a\}\seq& -\half [\etab,\etab]^\a\ \ ,\cr
\{{\bar Q},\Psi_\a\}\seq& [\Psi,\etab]_\a\ \ .\cr}\eqno(\EqQTRANS)$$
Now the Hamiltonians of TQFT's are given by [\RefWittenTQFT]
$$H\seq\half \{Q,{\bar Q}\}\ \ .\eqno(\EqHQQBAR)$$
Postulating the existence of a TQFT
for which the BRST and anti-BRST charges are as given in (\EqQ-\EqQTRANS),
we find that its Hamiltonian is
$$H\seq \half \int_\Sigma \big(\Psi^\a\Psi_\a \mi
\Psi^\a[\etab,\eta]_\a \pls \half [\etab,\eta]^2\big)\ \ .\eqno(\EqH)$$
This Hamiltonian was constructed starting from a phase space with
coordinates $(p,q)$ and extension $(p,q,\eta,\etab)$ along with a set
of first class, irreducible constraints $\Psi_\a(p,q)$.  Alternatively,
we could have started by writing down $H$ and then realizing that
there existed a BRST charge for which $H\in \image(Q)$.  In the
latter case, we would then require that our physical states be in
the $Q$-cohomology.  Subsequently, we would arrive at a physical
Hilbert space given by $\ker(\Psi_\a)$.

For the theory with
Hamiltonian (EqH) to be correctly labelled a TQFT, its
energy-momentum tensor should be $Q$-exact.  The most direct way to
check this is to construct the action.  This is beyond the scope of
this note.  At this stage, it is more fruitful to check for a
relations to canonical gravity. Let us now find a homomorphism
between the constraints and $\diffs$.

Consider a orientable manifold, $M$, of dimension $D=n+2$
with $n\geq0$.
Introduce a $n$-form \hbox{$B^A=dx^{a_1}\wedge\ldots\wedge
dx^{a_n}B_{a_1,\ldots ,a_n}{}^A$} which is valued in a
semi-simple Lie Group \LG.
Introduce the gauge field $A_a{}^A$ with field-strength
$F^A= dx^b\wedge dx^c F_{bc}{}^A$.  The
manifestly diffeomorphism invariant action constructed
out of these forms is [\RefHorowitz]
$$S\seq  \inv{2n!}\int_M Tr(B\wedge F)\ \ .\eqno(\EqSBF)$$
$S$ is also invariant under the $n$-form symmetry
\hbox{$B\to B+D\Theta$},
for some {\LG}-valued $(n-1)$-form $\Theta$. ($D$ is the
exterior covariant derivative).  We immediately see that
$B$ acts as a Lagrange multiplier which imposes $F=0$
as an equation of motion.  On the other hand, the $A$
equation of motion yields $DB=0$.  As we will canonically
quantize $S$, let us take $M$ to be of the form $M=\Sigma\times\IR$.
We then find the constraints
$$\cg \sp\equiv\sp D \dual{B}
 \sp\approx\sp 0\qquad {\rm and}\qquad \cp\sp\equiv\sp
 F\sp\approx\sp 0\ \ ,\eqno(\EqCons)$$
where $^*$ is the Hodge star operator on $\Sigma$.

We now consider the algebra of these constraints.
The canonical ETC between momenta and coordinates is given by
$$[\dual{B}_i{}^A(x),A^j{}_B(y)]\seq -i \d_B{}^A\d_i{}^j
\d_\Sigma(\vec{x}-\vec{y})\ \ ,\eqno(\EqETC)$$
where $i, j$, etc. are indices restricted to the $\Sigma$
hypersurface.  The Hamiltonian density is a sum of the
constraints, $\ch=Tr(A_0\cg +  \dual{B}_0{}^{ij} \cp_{ij})$.
It
is straightforward to show that the algebra of constraints is
$$
\eqalign{
[\cg_A(x),\cg_B(y)]\seq& i f_{AB}{}^C\cg_C(x)\dsig\ \ ,\cr
[\cg_A(x),\cp_{ijB}(y)]\seq & i f_{AB}{}^C \cp_{ijC}(x)\dsig\ \ ,\cr
[\cp_{ijA}(x),\cp_{klB}(y)]\seq &0\ \  .\cr}\eqno(\EqCA)$$
Gauge transformations with smearing functions $\Lambda^A$ are generated
by
$G[\Lambda]=-i\int_\Sigma \Lambda^A\cg_A$.
Likewise, the $n$-form symmetry is generated by
$P[\Theta]=i\int_\Sigma \Theta^A\cp_A$.

Now introduce a scalar function $M$ and a vector field $\vec{M}$,
then define the operators
$$\eqalign{
\IH[\vec{M}]\seq& i\int d^{n+1}y M^i{\rm Tr}\big(\dual{B}^j\cp_{ij}
\pls A_i\cg\big)\ \ ,\cr
\ihp[M]\seq& i\int d^{n+1}y M {\rm Tr}\big([\dual{B}^i,
\dual{B}^j]\cp_{ij}\big)\ \ ,\cr}\eqno(\EqIH)$$
These definitions are motivated by Ashtekar's [\RefAsh] approach to
non-perturbative quantum gravity.  The algebra of these operators is
found to be
$$\eqalign{
[\IH[\vec{M}],\IH[\vec{N}]]\seq& \IH[[\vec{M},\vec{N}]]\ \ ,\cr
[\IH[\vec{M}],\ihp[{N}]]\seq& \ihp[{\cal L}_{\vec M}N]\ \ ,\cr
[\ihp[{M}],\ihp[{N}]]\seq& -i\int d^{n+1}y (M\lrpd_iN){\rm Tr}
\big([[\dual{B}^i,\dual{B}^j],\dual{B}^k]\cp_{jk}\big)\ \ ,\cr}
\eqno(\EqFDIFF)$$
from the first line, we notice that $\IH[\vec M]$ is a homomorphism
from the
$\diffs$ Lie algebra into the constraint algebra of the $BF$ theory for
{\it arbitrary} \LG.
Although this set of three commutators does not close for general \LG,
it is still first class.  It is only known to be first class in GR
for rank one groups; then the last term simplifies and the
algebra closes but it is not a Lie Algebra.  To be precise, for
$\LG=SO(2,1)$, we find the full algebra to be
$$\eqalign{
[\IH[\vec{M}],\IH[\vec{N}]]\seq& \IH[[\vec{M},\vec{N}]]\ \ ,\cr
[\IH[\vec{M}],\ihp[{N}]]\seq& \ihp[{\cal L}_{\vec M}N]\ \ ,\cr
[\ihp[{M}],\ihp[{N}]]\seq&\IH[M\lrpd_iN]\pls G[(N\lrpd_iM) g^{ij}A_j]
\ \ ,\cr
[G[\Lambda],G[\Gamma]]\seq& G[[\Lambda,\Gamma]]\ \ ,\cr
[G[\Lambda],\IH[\vec{M}]]\seq& G[{\cal L}_{\vec M}\Lambda]\ \ ,\cr
[G[\Lambda],\ihp[M]]\seq& 0\ \ ,\cr}
\eqno(\EqDIFF)$$
where we have identified the inverse metric on the hypersurface:
$g^{ij}\equiv \dual{B^i}{}_A\dual{B}^{jA}$.  $\dual{B}^i{}_A$ is
allowed to be degenerate as we did not introduce its inverse.
These features are familiar from previous work in GR [\RefAsh].
It has been shown [\RefBengtson] that for D=3, $\IH[\vec{M}]\approx0$
and $\cp\approx0$ and  $\cg\approx0$ imply each other if
$\det(g^{ij})\neq0$;
\ie, if $g_{ij}$ is non-degenerate.  Unlike GR, we realize that we
could have identified the metric on $\Sigma$ before working out the
algebra (\EqDIFF).  From eqn. (\EqCons), we see that the constraints
may be solved by taking $\dual{B}_i{}^A$ to be the dreibein and
$A_i{}^A$ to be the Lorentz spin-connection restricted to $\Sigma$.
In particular, $\cg\approx0$, is the torsion free constraint familiar
from [\RefWittenCSG].

In D=3, the constraints (\EqCons) are irreducible.
Consequently, we can apply eqns. (\EqQ) -- (\EqH) without alteration.
To be precise, we can now claim that for $D=3$ and $\LG=SO(2,1)$,
$$\eqalign{
H\seq \int_\Sigma {\rm Tr}\big(
&\half D_i\dual{B}^{i}D_i\dual{B}^{j}\pls F_{ij}F^{ij} \cr
&-\sp D_i \dual{B}^{i}[\rhob,\rho]\mi \half D_i \dual{B}^{i}[\xib,\xi] \cr
&+\sp \inv{4} [\xib,\xi]^2 \pls \half [\rhob,\rho]^2 \pls \half
[\xib,\xi][\rhob,\rho]\big)\ \ ,\cr}\eqno(\EqHthree)$$
is the Hamiltonian of a TQFT whose constraints are homomorphic to
$\diffs$ with the metric $g_{ij}=\dual{B}_i{}^A\dual{B}_{ja}$ on the
hypersurfaces
at fixed time.  Here, in the notation of eqn. (\EqH),
$\eta^\a\in\{\xi^A,\rho^A\}$ with $\xi^A$ being the Gauss law constraint
ghost and $\rho^A$ the flat-connection constraint ghost.
The explicit form and action of $Q$ are
$$\eqalign{
Q\seq \int_\Sigma \big(\xi^A&\cg_A \pls \rho^A\cp_A \mi \half
\xib^A[\xi,\xi]_A \mi \rhob^A[\xi,\rho]_A\big)\ \ ,\cr
\{Q,\xi^A\}\seq&-\half[\xi,\xi]^A\ \ ,\cr
\{Q,\xib^A\}\seq& \cg^A \mi [\xib,\xi]^A\mi [\rhob,\rho]^A\ \ ,\cr
\{Q,\rho^A\}\seq& -[\xi,\rho]^A\ \ ,\cr
\{Q,\rhob^A\}\seq& \cp^A\mi [\rhob,\xi]^A\ \ ,\cr
[Q,\cg_A]\seq& [\cg,\xi]_A\pls [\cp,\rho]_A\ \ ,\cr
[Q,\cp_A]\seq& [\cp,\xi]_A\ \ ,\cr
[Q,\dual{B}^{iA}]\seq& i[\dual{B}^i,\xi]^A\pls iD^i\rho^A\ \ ,\cr
[Q,A^{iA}]\seq& iD^i\xi^A\ \ .\cr}\eqno(\EqQthree)$$
We note from that the last commutator that the gauge field does not
transform via a shift as is the case with  Topological Yang-Mills (TYM)
[\RefWittenTQFT].  In fact, the action of $Q$ on $A$ yields a Yang-Mills
gauge transformation.  Thus, although the Hamiltonian is from a TQFT
it is unclear whether or not it is from a TYM theory.

The constraints in two dimensional space-time are different in character
from those of higher dimensions.  Since $2$-forms do not exist on
one-dimensional manifolds, there is no $\cp\approx0$ constraint. The only
constraint is that of Gauss's Law, $\cg^A=D_1B^A\approx0$, where $B^A$ is
now a $0$-form.  As in [\RefChamWyl] we treat both Poincar{\' e} and
(anti-)de Sitter gravity by taking $\LG = SO(1,2)$ with arbitrary
cosmological constant, $\l$, so that the $SO(1,2)$ algebra is of the
form
$[P_a,P_b]=\l \e_{ab} J$ and $[J,P_a]= \e_{ab}P^b$.
The extended phase space includes the ghost coordinates $\th^a$ and
$\z$ along with the anti-ghosts $\thb^a$ and $\zb$. With these and
$M=S^1\times\IR$, the Hamiltonian of our TQFT reads
$$\eqalign{
H\seq\half \int_{S^1} \big(&\l D_1B^a D_1B_a \pls D_1BD_1B \pls \l
D_1B^a(\zb\th^b\mi\thb^b\z)\e_{ab}\cr
&-\sp\l D_1B\thb^a\th^b\e_{ab} \mi \l \zb\z\thb^a\th_a\pls \half\l^2
\thb^a\thb^b\z_a\z_b\big)\ \ .\cr}\eqno(\EqHtwo)$$
In these 1+1 dimensional cases it is unclear how to identify the metrics.

In conclusion, we have found a way to identify the constraints  and
phase spaces of General Relativity in $2+1$ and $1+1$ dimensions with
the constraints and phase spaces of theories whose Hamiltonian are of
the TQFT form. Metrics on the
two-dimensional hypersurfaces were also identified. The method
combines the BFV quantization of constrained systems along with the
BRST exactness of the  Hamiltonians of TQFT's.  It is expected that
generalization to four dimensions is possible.
\vfill\eject
\refs
\ref{\RefGrossMende}{D.J. Gross and P. F. Mende, \pl{197} (1987) 129;
\np{303} (1988) 407}
\ref{\RefWittenTQFT}{E. Witten, \cmp{117} (1988) 353}
\ref{\RefBrooksSpec}{R. Brooks, \np{325} (1989) 481}
\ref{\RefAchucarro}{A. Ach\'ucarro and P. Townsend,
\pl{180} (1986) 89}
\ref{\RefWittenCSG}{E. Witten, \np{311} (1988) 46}
\ref{\RefChamseddine}{A.H. Chamseddine, \pl{233} (1989) 291; \np{346}
(1990) 213}
\ref{\RefChamWyl}{A.H. Chamseddine and D. Wyler, \pl{228} (1989) 75;
\np{340} (1990) 595}
\ref{\RefAsh}{A. Ashtekar,
{\sl Non-perturbative Canonical
Quantum Gravity}, (World Scientific, Singapore, 1991);
C. Rovelli, \cqg{8}
(1991) 1613}
\ref{\RefBengtson}{I. Bengtsson, \pl{220} (1989) 51}
\ref{\RefRomano}{A. Ashtekar and J.D. Romano, \pl{229} (1989) 56}
\ref{\RefBirminghamRev}{For a review see D. Birmingham,
M. Blau, M. Rakowski and G.
Thompson, \pr{209} (1991) 129}
\ref{\RefHorowitz}{G. Horowitz, \cmp{125} (1989) 417}
\ref{\RefBFV1}{I.A. Batalin and G.A. Vilkovisky, \pl{69} (1977) 309;
I.A. Batalin and E.S. Fradkin, \pl{128} (1983) 303}

\bye